\documentclass{webofc}
\usepackage[varg]{txfonts}   
\newcommand{\PL}{{\fontfamily{cm}\sf\selectfont PL}}
\newcommand{\WSdeltas}{{\fontfamily{cm}\sf\selectfont WS$_{\delta({\bf s})}$}}
\newcommand{\WS}{{\fontfamily{cm}\sf\selectfont WS}}
\begin{document}
\title{Illuminating the impact-parameter dependence of UPC dijet photoproduction}

\author{\firstname{Kari~J.} \lastname{Eskola} \and \firstname{Vadim} \lastname{Guzey} \and \firstname{Ilkka} \lastname{Helenius} \and \firstname{Petja} \lastname{Paakkinen}\thanks{Speaker, \email{petja.k.m.paakkinen@jyu.fi}} \and \firstname{Hannu} \lastname{Paukkunen}
}

\institute{University of Jyväskylä, Department of Physics,
P.O. Box 35, 40014 University of Jyväskylä, Finland
\and Helsinki Institute of Physics,
P.O. Box 64, 00014 University of Helsinki, Finland
}

\abstract{%
  We present new NLO pQCD predictions for photoproduction of dijets in ultraperipheral PbPb collisions at 5.02 TeV with a realistic photon flux and up-to-date nuclear PDFs. Our calculation of the impact parameter dependence of the photon flux includes the effects of the nuclear form factor in the photon-emitting nucleus and the spatial dependence of nuclear PDFs of the target nucleus, which are estimated using the Wood-Saxon nuclear density profile. We show that a significant portion of the measured dijets at large $z_\gamma$ originate from events with impact parameters of the order of a few nuclear radii, and that the cross section predictions therefore become sensitive to the modelling of the nuclear geometry and photon flux close to the source nucleus.
}
\maketitle
\vspace{-0.2cm}
\section{Introduction}
\label{sec:intro}
In ultra-peripheral heavy-ion collisions (UPCs), two nuclei pass each other at an impact parameter larger than the sum of their radii, suppressing the hadronic strong interactions. Hard interactions of one nucleus with the electromagnetic field of the other are still allowed, providing a way to probe the nuclear contents through photon-induced processes~\cite{Bertulani:2005ru}. For example, dijet photoproduction has been promoted as a way to probe the nuclear PDFs~\cite{Strikman:2005yv,Guzey:2018dlm}. Previous NLO pQCD predictions for this process have been performed by using a point-like charge distribution for the nucleus to simplify the calculations. We show that this picture receives corrections due to geometrical effects originating from the finite size of the involved nuclei.

\section{Impact-parameter dependent factorization of dijet cross section}
\label{sec:xsec}
In this work, we follow the approach of Refs.~\cite{Baron:1993nk,Greiner:1994db,Krauss:1997vr}, where the photon-induced process is treated in the equivalent-photon approximation with explicit transverse-plane dependence and keeping track of the finite-size effects. For the dijet production we have
\begin{equation}
  \begin{split}
    {\rm d} \sigma^{AB \rightarrow A + {\rm dijet} + X} = \sum_{i,j,X'} \int & {\rm d}^2{\bf b} \, \Gamma_{AB}({\bf b}) \int {\rm d}^2{\bf r} \, f_{\gamma/A}(y,{\bf r}) \otimes f_{i/\gamma}(x_\gamma,Q^2) \\
    \otimes \int & {\rm d}^2{\bf s} \, f_{j/B}(x,Q^2,{\bf s}) \otimes {\rm d} \hat{\sigma}^{ij \rightarrow {\rm dijet} + X'} \delta({\bf r}\!-\!{\bf s}\!-\!{\bf b}).
  \end{split}
  \label{eq:xsec_full}
\end{equation}
Here, $\Gamma_{AB}({\bf b})$ gives the probability of no hadronic interaction between the involved nuclei at an impact parameter ${\bf b}$, $f_{\gamma/A}(y,{\bf r})$ is the number of photons at a transverse distance $|{\bf r}|$ from the center of the emitting nucleus $A$ carrying a fraction $y$ of the beam energy, and $f_{i/\gamma}(x_\gamma,Q^2)$ the density of partons $i \in \{q,\bar{q},g,\gamma\}$ with a momentum fraction $x_\gamma$ in the photon at the scale $Q^2$. Here, $f_{\gamma/\gamma}(x_\gamma,Q^2)=\delta(1-x_\gamma)$ accounts for the direct contribution. Finally, $f_{j/B}(x,Q^2,{\bf s})$ is the parton density for a parton $j \in \{q,\bar{q},g\}$ with a momentum fraction $x$ in the target nucleus $B$ at a transverse distance $|{\bf s}|$ from its center, and ${\rm d} \hat{\sigma}^{ij \rightarrow {\rm dijet} + X'}$ is the partonic cross section.

Assuming that we can factorize the target parton distribution further as $f_{j/B}(x,Q^2,{\bf s}) = \frac{1}{B} \, T_{B}({\bf s}) \times f_{j/B}(x,Q^2)$, where $f_{j/B}(x,Q^2)$ is the ordinary spatially averaged nuclear PDF and $T_{B}({\bf s})$ the nuclear thickness function, we get
\begin{equation}
  {\rm d} \sigma^{AB \rightarrow A + {\rm dijet} + X} = \sum_{i,j,X'} f_{\gamma/A}^{\rm eff}(y) \otimes f_{i/\gamma}(x_\gamma,Q^2) \otimes f_{j/B}(x,Q^2) \otimes {\rm d} \hat{\sigma}^{ij \rightarrow {\rm dijet} + X'}
  \label{eq:xsec_w_eff_flux}
\end{equation}
where
\begin{equation}
  f_{\gamma/A}^{\rm eff}(y) = \int {\rm d}^2{\bf r} \, f_{\gamma/A}^{\rm eff}(y,{\bf r}) = \frac{1}{B} \int {\rm d}^2{\bf r} \int {\rm d}^2{\bf s} \, f_{\gamma/A}(y,{\bf r}) \, T_{B}({\bf s}) \, \Gamma_{AB}({\bf r}\!-\!{\bf s})
  \label{eq:eff_flux}
\end{equation}
is an \emph{effective} photon flux containing now all the transverse-plane integrals (cf.~Ref.~\cite{ATLAS:2022cbd}).

\vspace{-0.2cm}
\section{Effective photon flux}
\label{sec:flux}

\begin{figure*}
\centering
\includegraphics[width=\textwidth]{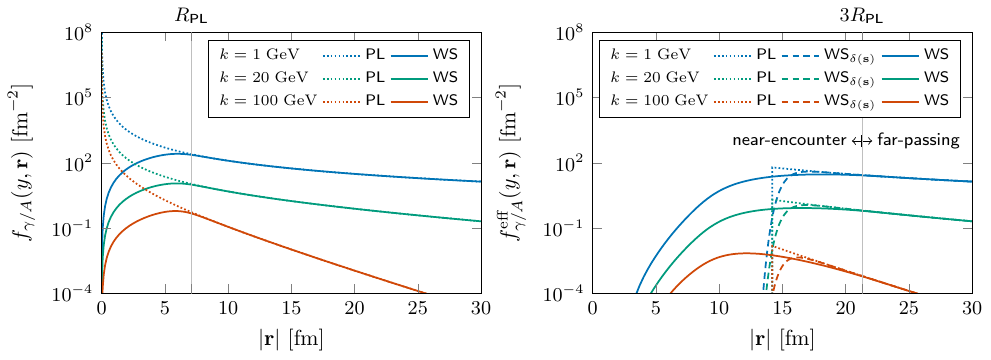}
\vspace{-0.5cm}
\caption{The bare (left) and effective (right) photon flux for different photon energies $k = y\sqrt{s_{\rm NN}}/2$ in PbPb collisions at $\sqrt{s_{\rm NN}} = 5.02\ {\rm TeV}$ under different approximations (see text).}
\vspace{-0.5cm}
\label{flux-r-dep}
\end{figure*}

Figure~\ref{flux-r-dep} shows the bare ($f_{\gamma/A}$) and effective ($f_{\gamma/A}^{\rm eff}$) photon flux in PbPb collisions at 5.02 TeV in three different approximations:
\begin{description}[\WS]
  \item[\PL] refers to the pointlike approximation, where the bare flux is taken to be that of a point source with a charge $Z$
  (see e.g.\ Ref.~\cite{Guzey:2018dlm}),
  \begin{equation}
    \textstyle
    f_{\gamma/A}^{\PL}(y,{\bf r}) = \frac{Z^2 \alpha_{\rm e.m.}}{\pi^2} m_p^2 y [ K_1^2(\xi) + \frac{1}{\gamma_L} K_0^2(\xi) ], \quad \xi = y m_p |{\bf r}|,
    \quad \gamma_L = \sqrt{s_{\rm NN}}/(2m_p),
  \end{equation} where $m_p$ is the proton mass. This bare \PL\ flux is shown as dotted lines in Figure~\ref{flux-r-dep} (left). The target nucleus is also taken as a pointlike object with $T_{B}({\bf s}) = B \delta({\bf s})$, and the requirement of no hadronic interaction between the nuclei is implemented with a simple cut in the impact parameter, $\Gamma_{AB}({\bf b}) = \theta(|{\bf b}| - b_{\rm min})$, where $b_{\rm min} = 2R_{\PL}$ is a tuneable parameter which we take as 14.2 fm in accordance with Ref.~\cite{Guzey:2018dlm}. Since we have ${\bf b} = {\bf r}$ in this approximation, this amounts to having a cut in the ${\bf r}$-dependent effective flux as shown in Figure~\ref{flux-r-dep} (right).
  \item[\WSdeltas] improves upon the previous one by deriving the bare flux through the nuclear form factor $F^{\WS}$ calculated from the Woods-Saxon nuclear density profile (cf.\ Ref.~\cite{Krauss:1997vr}),
  \begin{equation}
    \textstyle
    f_{\gamma/A}^{\WS}(y,{\bf r}) = \frac{Z^2 \alpha_{\rm e.m.}}{\pi^2} \frac{1}{y} \left| \int_0^\infty \frac{{\rm d}k_\perp k_\perp^2}{k_\perp^2 + (y m_p)^2} F^{\WS}(k_\perp^2 + (y m_p)^2)\,J_1(|{\bf r}|k_\perp) \right|^2,
  \end{equation}
  see the solid lines in Figure~\ref{flux-r-dep} (left). This only affects the distribution within the radius of the emitting nucleus $|{\bf r}| < R_{\PL}$, and thus a more important improvement comes from taking $\Gamma_{AB}({\bf b}) = \exp [-\,\sigma_{\rm NN}\,T_{AB}({\bf b})]$, where $T_{AB}({\bf b})$ is the nuclear overlap function calculated from the Woods-Saxon distribution, and $\sigma_{\rm NN} = 90\ {\rm mb}$ is the nucleon-nucleon total cross section at $\sqrt{s_{\rm NN}} = 5.02\ {\rm TeV}$. The effect is demonstrated in Figure~\ref{flux-r-dep} (right) with dashed lines. We still keep $T_{B}({\bf s}) = B \delta({\bf s})$ in Eq.~\eqref{eq:eff_flux} for this approximation.
  \item[\WS] is the most realistic approximation, where, on top of the previous one, also the target profile $T_{B}({\bf s})$ is calculated from the Woods-Saxon distribution. The resulting effective flux is shown as solid lines in Figure~\ref{flux-r-dep} (right), where we see that the additional integration over finite ${\bf s}$ gives a non-negligible contribution in the region $|{\bf r}| < 2R_{\PL}$ in contrast to the two other approximations. This effect becomes increasingly important for large photon energies where the bare flux is very steeply falling as a function of $|{\bf r}|$.
\end{description}

For the events with $|{\bf r}| > 3R_{\PL}$, i.e.\ `far-passing' nuclei, the PL approximation works fine, but producing high-$p_{\rm T}$ jets requires sufficient energy from the photon, which enhances sensitivity to the `near-encounter' $|{\bf r}| < 3R_{\PL}$ region, as we discuss next.

\vspace{-0.2cm}
\section{UPC dijet cross section in PbPb collisions at 5.02 TeV}
\label{sec:results}

\begin{figure*}
\centering
\includegraphics[width=\textwidth]{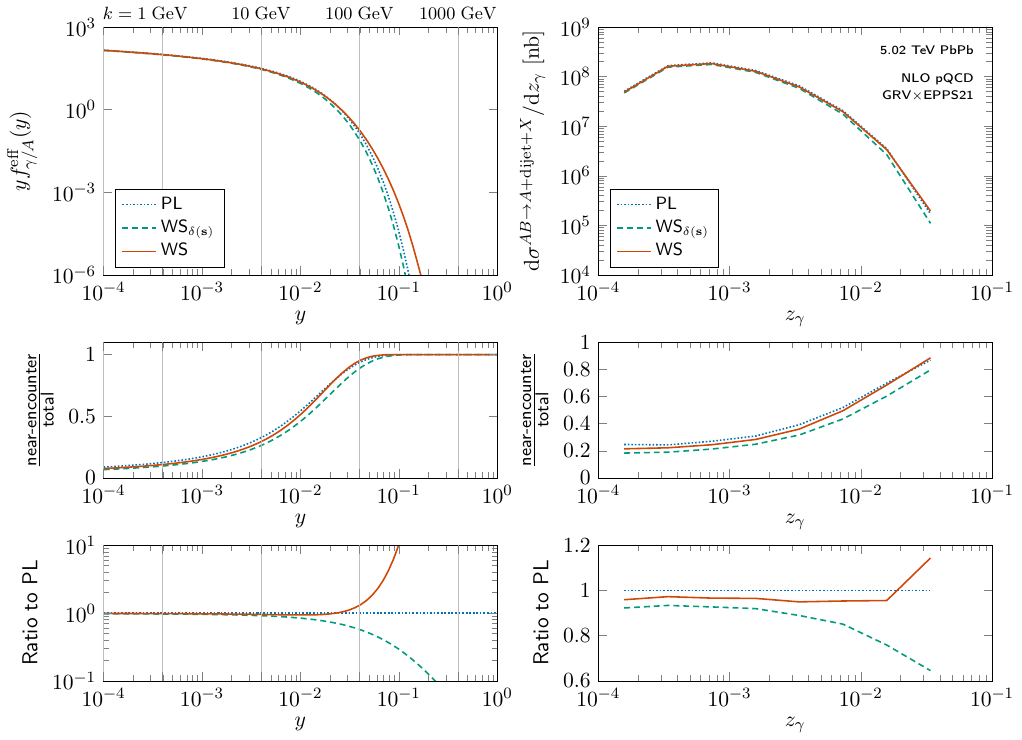}
\vspace{-0.5cm}
\caption{The effective flux (left) and UPC dijet cross section (right) in PbPb collisions at $\sqrt{s_{\rm NN}} = 5.02\ {\rm TeV}$ under different approximations (see text).}
\vspace{-0.5cm}
\label{flux-and-xsec}
\end{figure*}

We compare in Figure~\ref{flux-and-xsec} the effective flux integrated over the transverse-plane variables (left) and the inclusive dijet cross section as a function of $z_\gamma = M_{\rm jets} \exp(y_{\rm jets}) / \sqrt{s_{\rm NN}}$ (right) in PbPb collisions at $\sqrt{s_{\rm NN}} = 5.02\ {\rm TeV}$, where $M_{\rm jets}$, $y_{\rm jets}$ are the invariant mass and rapidity of the jet system. Jets are defined using the anti-$k_{\rm T}$ algorithm with $R = 0.4$ and are required to satisfy $|\eta_{\rm jet}| < 4.4$ with at least one of them having $p_{\rm T,jet} > 20\ {\rm GeV}$ and the rest $p_{\rm T,jet} > 15\ {\rm GeV}$ with $M_{\rm jets} > 35\ {\rm GeV}$, in accordance with Ref.~\cite{ATLAS:2017kwa}. At leading order, this observable would resolve to $z_\gamma = y x_\gamma$. Here, we calculate the cross section in NLO pQCD with the Frixione \& Ridolfi jet photoproduction code~\cite{Frixione:1997ks} using the EPPS21 nuclear~\cite{Eskola:2021nhw} and the GRV photon PDFs~\cite{Gluck:1991jc}.

The upper panels of the figure show the absolute effective flux and cross section, whereas in the middle panels we show the fraction of flux and cross section coming from the `near-encounter' $|{\bf r}| < 3R_{\PL}$ region with respect to that integrated over the full phase space. This fraction is small for low-energy photons, but increases with growing energy, reaching unity at $y \approx 0.1$. Consequently, the fraction of the cross section coming from `near-encounter' events rises from 0.2 to 0.8 between the lowest and highest $z_\gamma$ bin. This indicates that the UPC dijets with large $z_\gamma$ are potentially sensitive to the modelling of collision geometry. Indeed, as we plot in the lower panels, the flux in \WSdeltas\ and \WS\ deviate from the \PL\ approximation for high photon energies, causing a 40\% suppression in \WSdeltas\ compared to \PL\ in the cross section in the largest $z_\gamma$ bin. The \PL\ calculation does a rather good job in approximating the full \WS\ one with some 10\% difference at maximum, which is of the order of the expected nuclear-PDF effects~\cite{Guzey:2018dlm} and thus also important, but the large difference between \WS\ and \WSdeltas\ shows that taking into account the full collision geometry with finite ${\bf s}$ is important for an accurate physical interpretation of this observable.

\vspace{-0.2cm}
\section{Conclusions}
\label{sec:concl}

Requiring an energetic photon in the initial state, a significant part of the UPC dijet cross section at large $z_\gamma$ comes from configurations where the nuclei pass each other at relatively small impact parameters. We have shown that this leads to a sensitivity to the nuclear transverse profile with a significant effect in the largest studied $z_\gamma$ bins. In the above, we have assumed that one can factorize $f_{j/B}(x,Q^2,{\bf s}) = \frac{1}{B} \, T_{B}({\bf s}) \times f_{j/B}(x,Q^2)$, but this is of course a simplification and one should use impact-parameter dependent nuclear PDFs~\cite{Frankfurt:2011cs,Helenius:2012wd} instead. Furthermore, we have neglected the possibility of electromagnetic breakup through Coulomb excitations. These matters will be discussed in an upcoming publication~\cite{inprep}.

\vspace{0.2cm}
\begin{acknowledgement}%
  This research was funded through the Academy of Finland project No.~330448, as a part of the Center of Excellence in Quark Matter of the Academy of Finland (projects No.~346325 and No.~346326) and as a part of the European Research Council project ERC-2018-ADG-835105 YoctoLHC. We acknowledge computing resources from the Finnish IT Center for Science (CSC), utilised under the project jyy2580.
\end{acknowledgement}

\end{document}